\documentclass{jaa}

\usepackage{graphicx}
\usepackage{url}
\usepackage{xcolor}
\usepackage{array}
\usepackage[authoryear]{natbib}

\newcommand{\aap}{    {\it A \& A}}

\newcommand{\aj}{     {\it AJ}}
\newcommand{\apj}{    {\it ApJ}}
\newcommand{\apjl}{   {\it ApJL}}

\newcommand{\mnras}{  {\it MNRAS}}
\newcommand{\nat}{    {\it Nature}}

\begin{document}\sloppy

\title{UVIT/ASTROSAT studies of Blue Straggler stars and post-mass transfer systems in star clusters: Detection of one more blue lurker in M67}


\author{Annapurni Subramaniam\textsuperscript{1,*}, Sindhu Pandey\textsuperscript{1}, Vikrant V. Jadhav\textsuperscript{1,2} and Snehalata Sahu\textsuperscript{1}}
\affilOne{\textsuperscript{1}Indian Institute of Astrophysics, Sarjapur Road, Koramangala, Bangalore, India.\\}
\affilTwo{\textsuperscript{2}Joint Astronomy Programme and Department of Physics, Indian Institute of Science, Bangalore, India.
}


\twocolumn[{

\maketitle

\corres{purni@iiap.res.in}

\msinfo{08 September 2020}{08 September 2020}

\begin{abstract}
The blue straggler stars (BSSs) are main-sequence (MS) stars, which have evaded stellar evolution by acquiring mass while on the MS. 
The detection of extremely low mass (ELM) white dwarf (WD) companions to two BSSs and one yellow straggler star (YSS) from our earlier study using UVIT/\textit{ASTROSAT}, as well as WD companions to  main-sequence stars (known as blue lurkers) suggest a good fraction of post-mass transfer binaries in M67. Using deeper UVIT observations, here we report the detection of another blue lurker in M67, with an ELM WD companion. The post-mass transfer systems with the presence of ELM WDs, including BSSs, are formed from Case A/B mass transfer and are unlikely to show any difference in surface abundances. We find  a correlation between the temperature of the WD and the $v\ sin i$ of the BSSs. We also find that the progenitors of the massive WDs are likely to belong to the hot and luminous group of BSSs in M67. The only detected BSS+WD system by UVIT in the globular cluster NGC 5466, has a normal WD and suggests that open cluster like environment might be present in the outskirts of low density globular clusters. 
\end{abstract}


}]


\doinum{12.3456/s78910-011-012-3}
\artcitid{\#\#\#\#}
\volnum{000}
\year{0000}
\pgrange{1--}
\setcounter{page}{1}
\lp{1}

\section{Introduction}
The blue straggler stars (BSSs) are brighter and bluer than the main sequence (MS) turnoff in the colour-magnitude diagram (CMD) of a star cluster, a region from which most of the stars of same mass and age have evolved. \cite{1953Sandage} discovered these stars in the globular cluster (GC) M3. They stay on the MS longer than the time estimated by the standard theory of stellar evolution.  
The formation pathways and evolution of these stars are not well understood as they defy standard single stellar evolutionary theory. They seem to be rejuvenated stars, and gained mass as a consequence of some uncommon process. There have been various possible explanation for the BSSs, with three widely accepted mechanism in the literature: (i) stellar collision of two or more stars in dense environment \citep{Hills1976}, (ii) mass transfer (MT) through Roche lobe over flow (RLOF) in a binary system \citep{McCrea1964} and (iii) MT resulting in a coalescence of the inner binary stars in hierarchical triples (\citealp{Perets2009}; \citealp{Naoz2014}). Though these mechanism have been strongly supported with several evidence, there are some BSSs that cannot be explained by these models \citep{Cannon2015ebss}.
\par
\begin{figure*}
    \centering
    \includegraphics[width=0.7\textwidth]{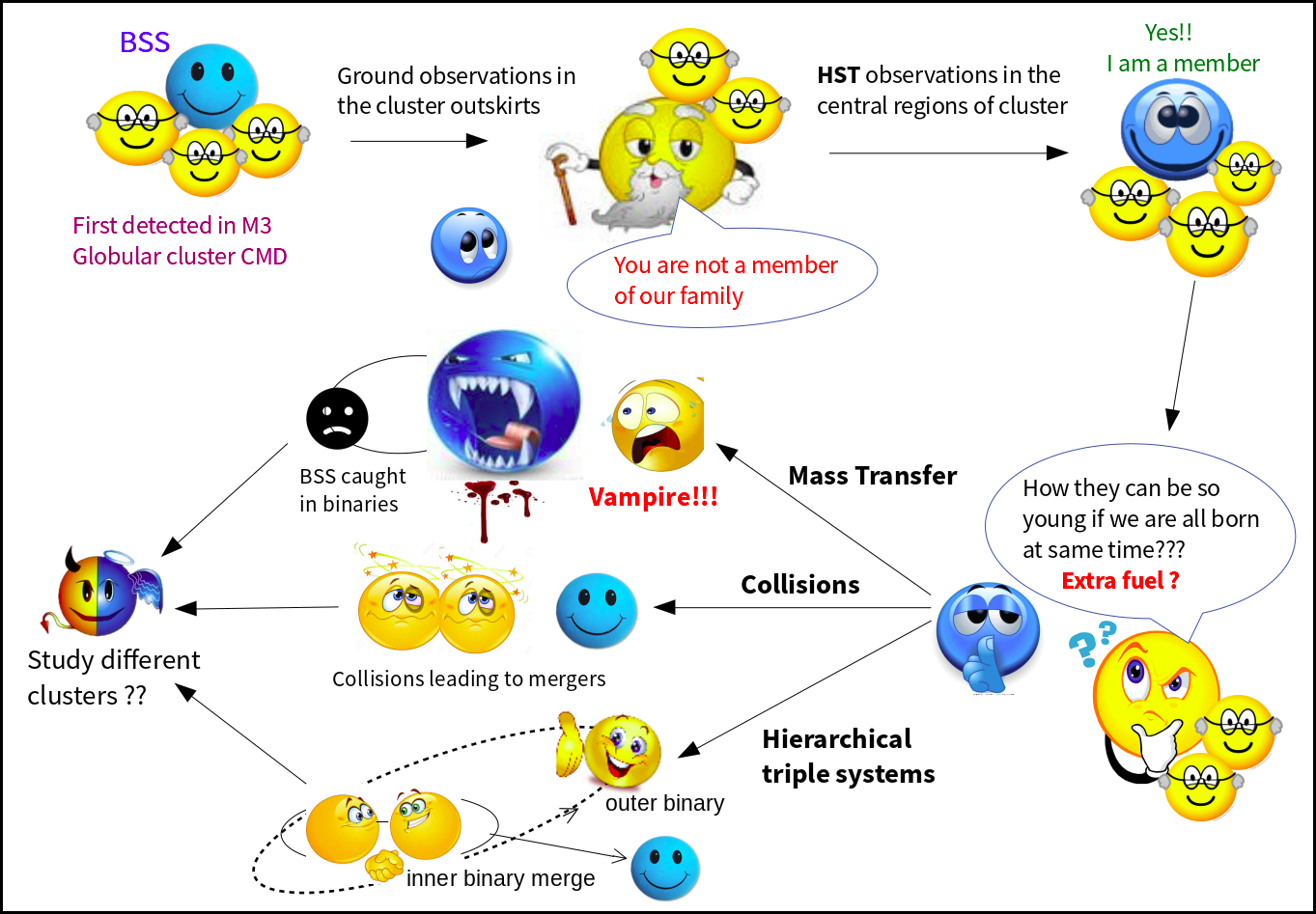}
    \caption{Formation pathways of BSS - a graphical illustration. BSSs were first detected in GC M3 and were initially discarded as possible non-members. Using HST observations and proper motion estimations, they were confirmed as members. This opened up the question of how the old clusters harbor these younger stars which still have enough hydrogen fuel to burn in their core. Three possible ways to gain mass are demonstrated  (1) the BSSs might have gained the extra fuel by sucking mass from their companion star (Vampire-like stars), (2) stellar collisions and (3) hierarchical triple systems.  (Image credits: Snehalata Sahu).}
    \label{fig:bss_cartoon_big}
\end{figure*}

A large number of BSSs are found in clusters, about 1887 BSS candidates in 427 open clusters (OCs) are catalogued \citep{Ahumada2007}. M67 with its several studies on BSSs has been the primary laboratory (\citealp{Mathieu1986}; \citealp{Mathys1991}; \citealp{Gilliland1992}; \citealp{Landsman1998}; \citealp{Deng1999}; \citealp{Liu2008_BSS}; \citealp{Bertelli2018}) while NGC 188 which has 20 BSSs in the cluster has been developed as a new laboratory (\citealp{Mathieu2009Natur}; \citealp{Geller2011Natur}; \citealp{Geller2013}; \citealp{Mathieu2015ebss}; \citealp{Gosnell2015}) to understand the properties of BSSs found in OCs. \cite{Gosnell2015} detected white dwarf (WD) companions to BSSs in NGC 188 and indicated that the BSSs in the cluster were possibly formed through MT. The likelihood of a single-single star collision in a less dense environment such as OC is prohibitively small. For a direct collision to occur, a dynamical environment involving encounters of single and or binary systems is essential. The MT could occur through either of the Case A, B or C depending on the evolutionary state of the donor. A helium WD companion is expected in a Case B MT, whereas CO WD in Case C MT. Case A MT, which generally happens in tight binaries through common envelope evolution, results in a merged product. The BSSs in such cases may be a product of two MS stars, which might have come in contact through a stellar evolutionary process. In case a stable MT happens, sufficient mass to the secondary could be added to transform it into a BSS or due to the loss of orbital angular momentum via winds could lead to a tight binary system to form a merger product. A representation of the various formation pathways for the formation of BSSs is shown in Figure \ref{fig:bss_cartoon_big}. This is a graphic illustration to help understand the various formation pathways.

\par
 
\par
The theoretical model of BSSs in an OC environment predicts that they are likely formed through MT or merger. The binary frequency observed among the BSSs in OCs are high with most of them being single lined spectroscopic binaries (\citealp{Mathieu2009Natur}; \citealp{Geller2015}). The stellar rotational velocities of the BSSs are higher, in some cases $v\ sin i$ $\geq$ 50 km s$^{-1}$. The observation suggests that a few BSSs in NGC 188 and M67 are rotating more rapidly than the normal MS stars of similar effective temperature. The rapid rotation may be a clue for a recent stellar interaction, including a MT. The observational properties of a few BSSs in OCs have shown that they are 0.2 -- 0.8 M$_\odot$ more massive than the MSTO, suggesting that there could be multiple ways to gain mass.

Old OCs are also excellent test beds to study the properties of WDs \citep{Kalirai2010}.
In general, WDs detected in OCs are considered as the end products of single star evolution. Therefore, the mass of a WD that is formed recently in OCs is formed from a progenitor with the MS turn-off (MSTO) mass of the cluster. \citet{Williams2018} detected $\sim$ 50 WD candidates in M67 and estimated their mass and spectral type, where many WDs required a progenitor more massive than a single star at MSTO of M67. Hence, they concluded  that these high mass WDs detected are likely to be evolved from BSSs.  Similarly, \citet{Sindhu2019} detected WDs with mass $<$0.3 M$_\odot$ as a companion to a BSS in M67.  As the single star evolution takes a lot more time than the age of the universe to form such extremely low-mass WDs (ELMs),  presence of a binary system is needed for their formation. They must have undergone significant amount of mass loss during their evolution in short period binary systems \citep{Brown2010} and have not been able to ignite helium in their cores. 
 ELM WDs of mass up to 0.1 \(M_\odot\) were found to be part of contact binaries \citep{Marsh1995, Benvenuto2005}, whose progenitors transferred mass to the secondary, due to RLOF. 

Ultra-Violet (UV) images are very effective in identifying BSS binaries with a hot companion as they show excess emission in the UV, which in general, is not expected from BSSs alone. Based on the FUV spectroscopy and spectral energy distributions (SEDs) of 48 blue objects in 47 Tuc obtained with Hubble Space Telescope (HST), \cite{Knigge2008} discovered several interesting binary objects which also includes one BSS-WD binary in the cluster. \cite{Gosnell2014, Gosnell2015} detected WD companions to seven BSSs in the OC NGC 188 based on Far-UV (F140LP, F150LP, and F165LP) observations with the HST. \cite{Subramaniam2016} detected  a hot companion (post-AGB/HB) to a BSS in NGC 188 using UVIT data on \textit{ASTROSAT}, thus showing the importance of UV observations of BSSs.

We have detected WD companions to BSSs in the OC M67 \citep{Sindhu2019, Jadhav2019, Sindhu2020IAU}, and in the outskirts of the GC NGC 5466 \citep{sahu_2019}, using the UVIT images. \cite{Leiner2019} identified a fast rotating MS star and suggested that it could be a post-MT product. As this star is located on the MS, rather than the normal BSSs, which are brighter and bluer the MS tip, they named it a blue-lurker (BL). \cite{Jadhav2019}, detected the presence of WD companions to two MS stars in M67, which can be classified as BLs. One has a confirmed post-MT nature, as its companion is a ELM WD. Detecting such BL stars, which are basically BSSs in the MS, is not a trivial task. \cite{Leiner2019} used stellar rotation derived from Kepler/K2 data, whereas \cite{Jadhav2019} used multi-wavelength SED analysis with UVIT data playing a major role.

Here we present the confirmation of a WD companion to another MS star in M67, using a deeper observation of the M67 using UVIT. 
We also present a correlation between the rotation of the BSSs, BLs and the temperature of the companion WD. We also place the BSSs, post-MT systems and the progenitors of the massive WDs to understand the landscape of these enigmatic stars in M67. 



\section{Data and methods}
We have observed several star clusters using the \textit{ASTROSAT}, in the last 5 years. The open and globular clusters are imaged with the UVIT, in the Far-UV and Near-UV channels. After the failure of the NUV channel, the FUV channel is used to obtain images in multiple filters. 

The M67 cluster was observed on 23 April 2017 as part of the G07 cycle. These observations were carried out in 3 filters of the FUV channel, and the data from NUV channel were not available due to detector related issues. The analysis and results presented in \cite{Jadhav2019}, indicated that deeper observations are needed to confirm the detection of a WD companion to several MS stars. The follow up deeper observations were carried out on 19 December 2018 in two filters of the UVIT. The photometry and the CMDs are presented in Jadhav \textit{et al}. (2020, submitted). 




\section{Detection of another blue lurker}
As many BSSs are formed via stellar interactions, there is a chance that the BSSs have a binary companion. This companion can be the outer component of a triple system, or a remnant of a post MT donor star. Analysing the SEDs of such systems can reveal the parameters of the multiple components. The details of SED fitting method are presented in \citet{Jadhav2019}. As demonstrated by them, a binary system with sub-luminous hot component can be easily identified and characterised using UV to IR SED. 

\citet{Jadhav2019} detected ELM WD companions to WOCS2007, WOCS3001 and WOCS6006, hence these are post-MT systems. WOCS2007 is a known BSS, 
WOCS3001 and WOCS6006 are BL stars.  
 A few more objects required further observations to confirm the presence of hotter companions. M67 is therefore likely to have a relatively large number of post-MT systems, among the BSSs as well as the MS. Here we have performed the SED analysis of WOCS11005, by including the deeper UVIT data of M67. 

WOCS11005 is likely to be a single star as radial velocity variation was not observed \citep{Geller2015}. It lies near the main-sequence turn-off in the optical CMD and near the beginning of the BSS sequence in the UV-optical CMD. \citet{Melo2001} estimated a slow rotation of $v\ sin i=4.9\ km\ s^{-1}$.

\begin{figure}
    \centering
    \includegraphics[width=0.45\textwidth]{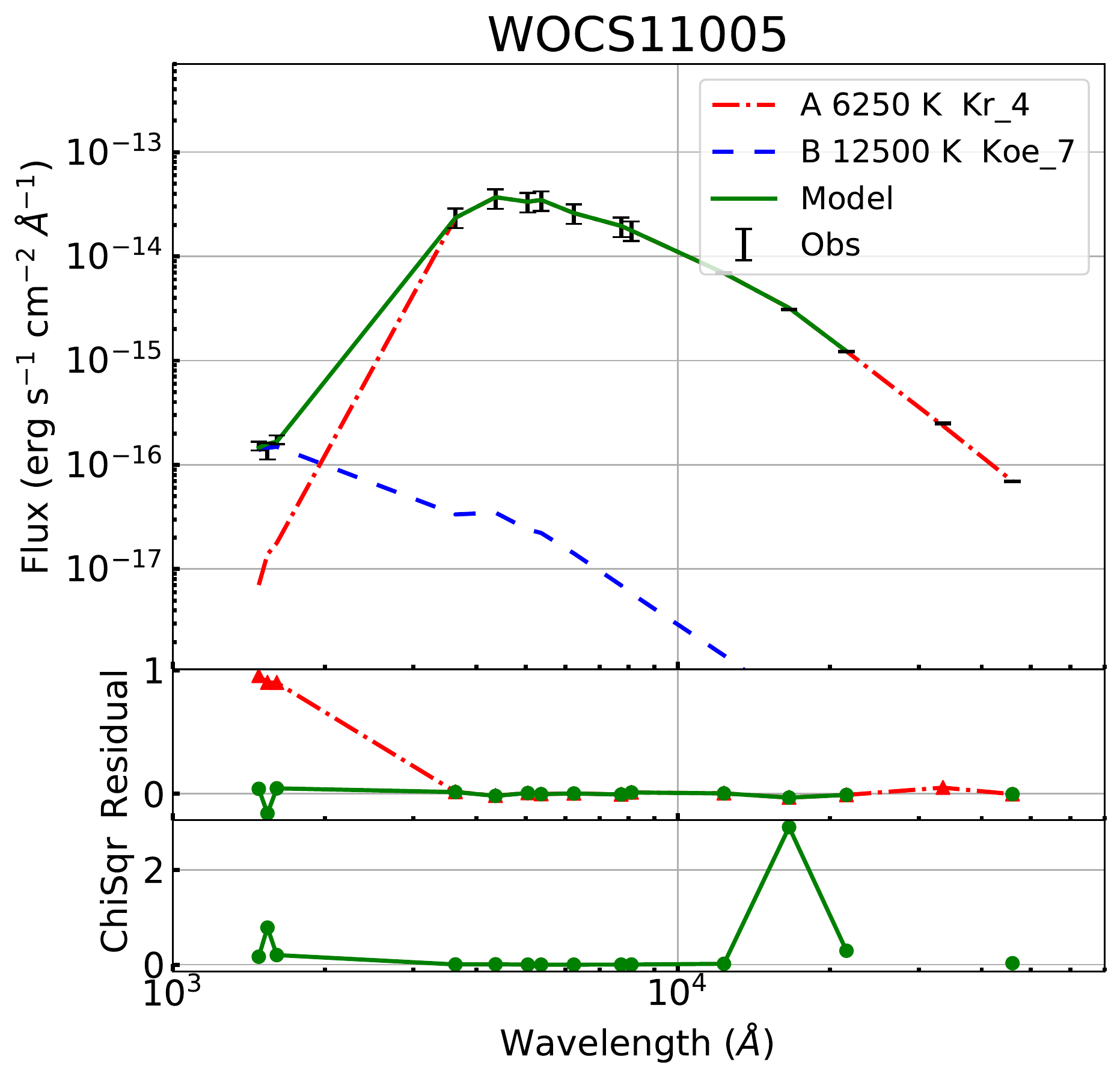}
    \caption{SED of M67 star `WOCS11005'. The top panel shows the cooler component (A, red dot-dashed), hotter component (B, blue dashed) and combined (green line) SED. Middle panel shows the fractional residual and bottom panel shows the individual $\chi^2$ values. The legend notes the $T_{eff}$ of the fit, the model used (Kr: Kurucz, Koe: Koester WD models) and the log $g$. }
    \label{fig:sed_M67}
\end{figure}

The source was detected in one UVIT filter (F148W) earlier. The {\it GALEX}  FUV flux is more or less consistent with the F148W flux from the UVIT and provides support to the UVIT detection. The new observations not only confirms the earlier F148W detection, but also estimates a very similar magnitude. The SED is shown in figure 2, where the 3 FUV data points consistent with one another, confirm the presence of a hot companion. The temperature  is estimated to be 12500K, slightly higher than that estimated by \citet{Jadhav2019}. The parameters of hotter companion lie within the predicted values of WD models. The age and mass of the WD are 1.29$\pm$1.16 Gyr and 0.27$\pm$0.01 M$_\odot$ respectively. 
The absence of detection in X-ray suggests minimal chromospheric activity and therefore the source of the high UV flux could be a possible WD. We therefore confirm the detection of a WD companion to WOCS11005. The companion is an ELM WD, suggesting that WOCS11005 is indeed a post-MT system. As this star is not located in the BS region, this is yet another BL star, but without rapid rotation.

\section{Relation between rotation and WD temperature}

\begin{figure}
    \centering
    \includegraphics[width=0.45\textwidth]{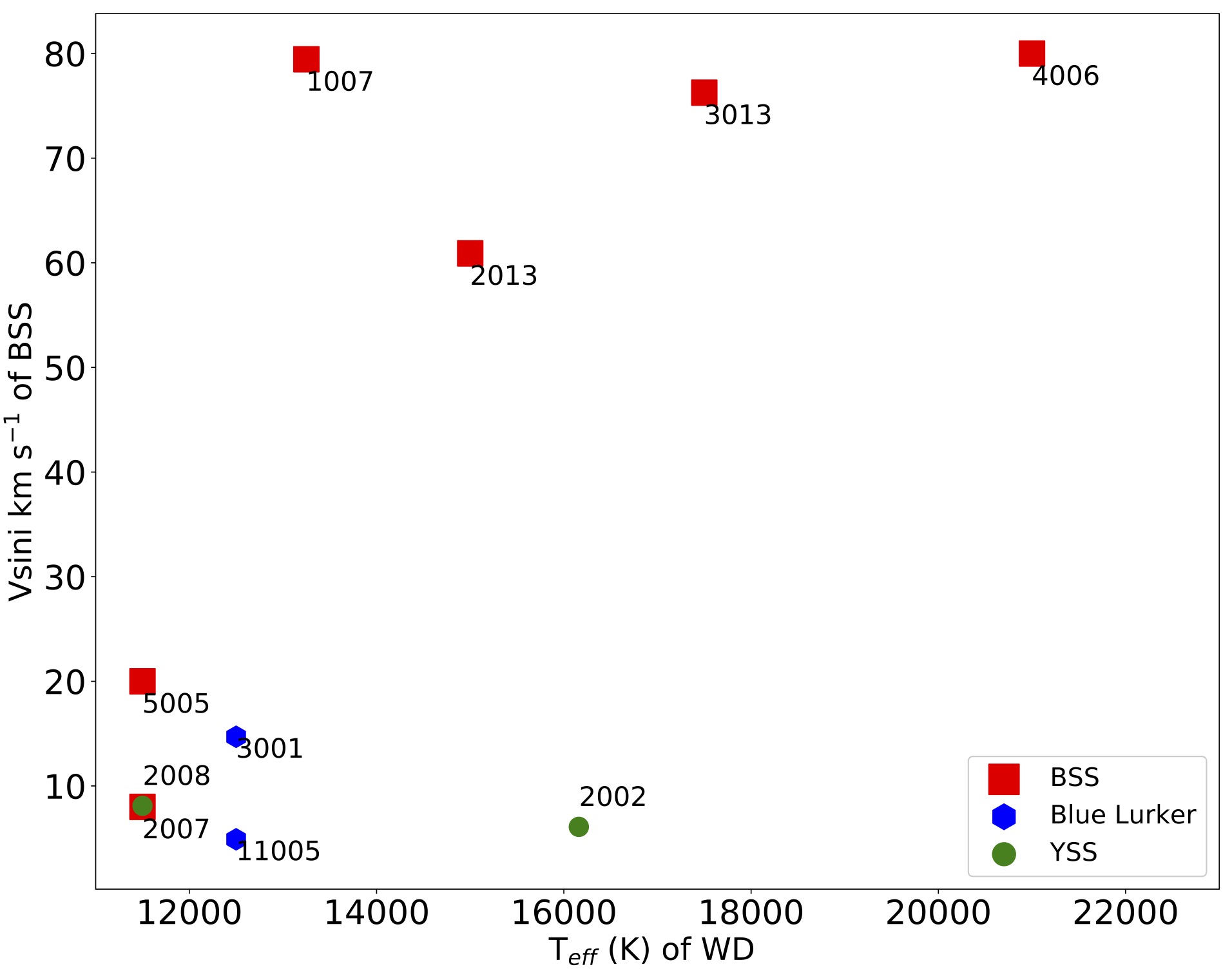}
    \caption{The rotational velocity of the BSSs/blue lurkers/YSSs against the temperature of the companion WD.}
    \label{fig:rotation}
\end{figure}

\cite{Leiner2019} suggested that stellar interactions, including MT can induce a spin-up, which enables identification of blue-straggler-type objects within the MS by identifying stars with fast rotation rates. They have put together a list of potential BLs. We analysed SEDs of 18 BSS and found that 5 BSS (WOCS1007, WOCS2013, WOCS3013, WOCS4006 and WOCS5005) have a hotter companion (\cite{Sindhu2020IAU} and Sindhu \textit{et al.} 2020 in prep.). In Fig.~\ref{fig:rotation}  (Table~\ref{tab:BSS_Vsini_T_WD} lists the parameters and classification), we have shown the plot of temperature of the WD companion against the $v\ sin i$ of all the  (BSS/BL/YSS)+WD systems detected so far in M67. It can be seen that there is a moderate correlation between the temperature of the WD and the rotation of the BSSs.  
The caveat is that this method will exclude systems which are pole on, and those with MT happened a while ago. WOCS3001 is the BL common with \cite{Leiner2019}  and \citet{Jadhav2019}. The rapid rotation of WOCS1007, WOCS2013, WOCS3013 and WOCS4006 suggest that they have undergone stellar interactions.

\begin{table}
     \caption{The rotational velocities and companion WD temperatures of BSSs/Blue lurkers/YSSs. References: (M89: \cite{Manteiga1989}, LM96: \cite{Latham1996}, L97: \cite{Landsman1997S1040}, V99: \cite{vandenberg1999}, M01: \cite{Melo2001}, M18: \cite{Motta2018}, L19:\cite{Leiner2019}, S19: \cite{Sindhu201} J19: \cite{Jadhav2019}, S20: \cite{Sindhu2020IAU}, Sp: Sindhu \textit{et al.} in prep)}

    \resizebox{0.48\textwidth}{!}{
    \begin{tabular}{cccr}
  \hline
    WOCS ID     & $v\ sin i$ & T$_{eff\ WD}$ & Comment\\
                & (km s$^{-1}$)  & (K) & \\ \hline
      1007   & 79.45 [M18] & 13250 [S19]& BSS\\
      2007 & 8.0 [M18]  & 11500 [J19]& BSS\\
      2013 & 60.90 [M18] & 15000  [Sp]& BSS\\
      3013 & 76.26 [M18]  & 17500  [Sp]& BSS\\
      4006 & 80 [M89]&  21000  [Sp]& BSS\\
      5005 & 20 [LM96] & 11500 [S20]& BSS\\
      11005& 4.9 [M01] & 12500  & BL\\
      3001 & 14.7 [L19] & 12500 [J19] & BL\\
      2008 & 8.1 [V99]  & 11500 [J19] & YSS\\
      2002& 6.09 [M18]  & 16160 [L97] & YSS\\
      \hline
    \end{tabular}
    }
    \label{tab:BSS_Vsini_T_WD}
\end{table}

\section {H-R diagram of BSS and massive WD progenitors}
The heavy WDs present in the M67 are likely formed from a BSS progenitor. \citet{Jadhav2019} derived the masses, luminosity and temperatures of the detected WDs. The mass of progenitor can be estimated by the empirical correlation found in \citet{Cummings2018}. We used their eq. (4) to estimate the progenitor masses of three heaviest WDs in M67. We estimated the temperature and luminosity of the progenitor by comparing the masses to BaSTI model \citep{Pietrinferni2004, Cordier2007}.

\begin{figure}
    \centering
    \includegraphics[width=0.4\textwidth]{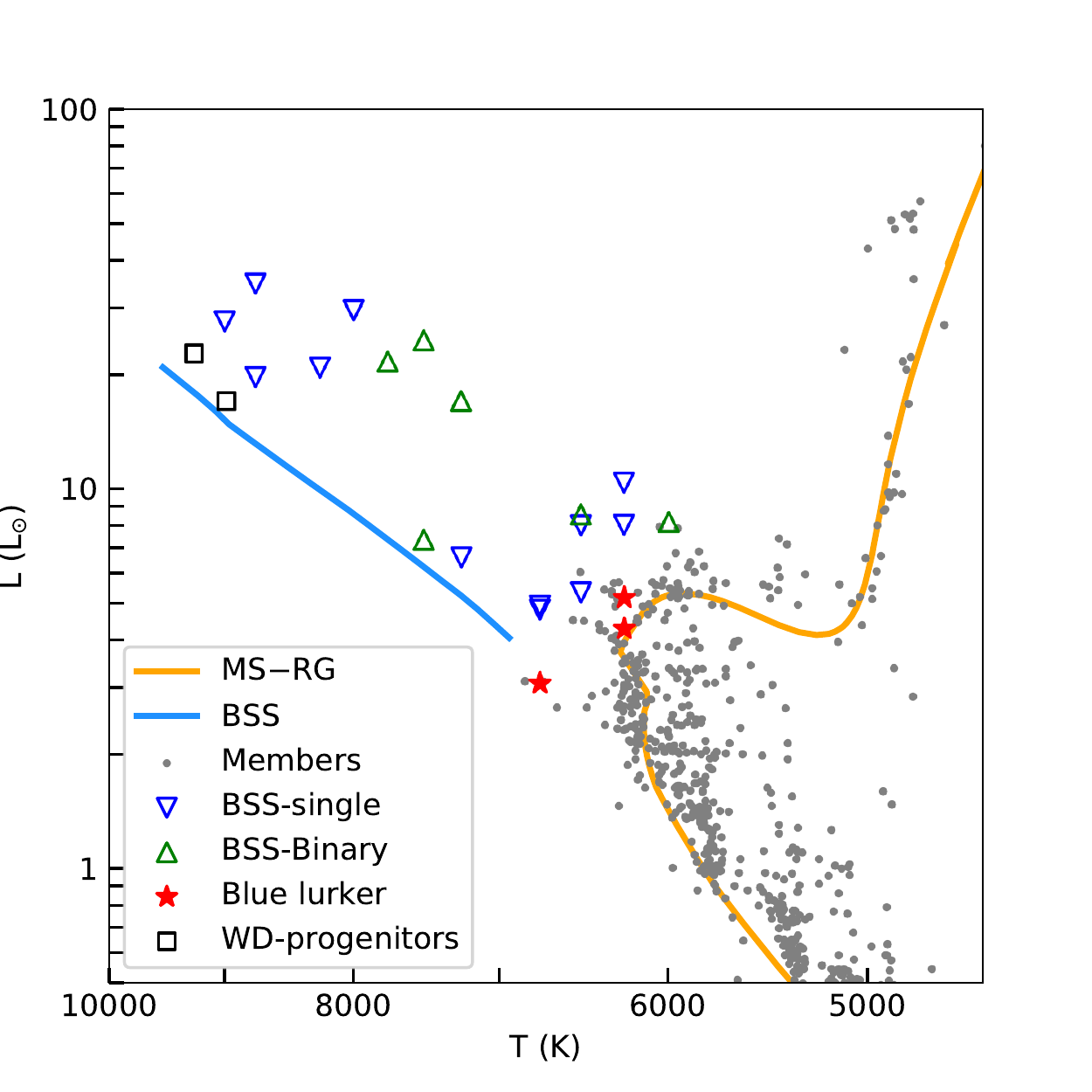}
    \caption{H--R diagram of M67 with isochrones generated with FSPS. The grey dots: Gaia DR2 T$_{eff}$ and luminosity of non-BSS members of M67. Blue inverted triangles: single BSS, green triangles: BSS components of BSS+WD systems, red stars: blue lurker components of BL+WD systems. Black squares: likely location of progenitors of the heaviest WDs in the cluster.}
    \label{fig:cmd_M67}
\end{figure}

We created the HR diagram of M67 members, including BSSs, blue lurkers and added the progenitors of the two massive WDs parameterised by \citet{Jadhav2019}.
Fig.~\ref{fig:cmd_M67} shows the BSSs which are single, binaries and post-MT systems.
As detailed in \cite{Sindhu2018}, the BSSs in M67 shows a clear demarcation with respect to their luminosity. The BSSs with Luminosity $L/L_\odot > 10.0$ form a brighter group. The location of the progenitors of the massive WDs are found very close to the luminous BSSs, adding to the group of 5 luminous single BSSs. This is an indication that the bright group of BSSs have been forming in the past as well. It is still not clear what makes these BSSs very luminous and what is their dominant formation pathway. 
Three blue lurkers are located close to the MSTO. This might indicate that these post-MT systems, are more likely to produce a relatively low luminous BSSs. These BLs will be the future BSSs of M67. In fact, figure Fig.~\ref{fig:cmd_M67}, depicts the current and some of the past and future BSSs in M67. 

\section{BSS+WD system in the Globular cluster NGC 5466}
The outskirts of GCs could host stable binaries and these might undergo MT, similar to those found in OCs. \cite{sahu_2019} found such a evidence and detected a BSS+WD system, that is confirmed to be a kinematic member. The hot companion is found to be a normal mass WD, and the BSS is expected to have gained mass from the envelope of the WD.  The WD is found be hot, with T$_{eff}$ = 32,000K, suggesting a recent MT. As this system is found in the outskirts of the low density cluster, this study confirms the presence of MT pathway of BSS formation in GCs. As several BSSs are detected in many GCs using FUV images of UVIT, there might be more such systems in GCs.

\section{Discussion}
The WDs companions found in the BSSs in M67 using UVIT images are found to have either a normal mass WD or an ELM WD. The normal mass WDs suggest that the core of the progenitor has evolved as a single star. The BSS could have gained mass of the envelope at the later stages of the evolution of the donor star. If the envelope had processed material, then the BSS might show the presence of chemical enhancement. On the other hand, BSSs with ELM WD companions are unlikely to show any chemical enhancement, as the MT must have happened before the core size of the donor has increased beyond 0.2 M$_\odot$. Therefore the MT is likely to have started near the sub-giant branch (Case A/B MT), resulting in the transfer of pristine envelope material. \cite{Motta2018} studied the chemical composition of 3 BSS candidates and two evolved BSSs in M67 and did not find any enhancement in Carbon. This might not be the case in other OCs.  The masses of WD companions to the BSSs in NGC 188, is found to peak at 0.5M$_\odot$ \citep{Geller2011Natur}. Our UVIT study is expected to throw light on the BSS formation pathways in other OCs.

As we detect a good number of ELM WD companions in M67, we suggest that the case A/B MT is one of the dominant MT mechanisms in this cluster. The UVIT studies have contributed to the detection of such systems in M67. In fact, two of the three BLs are found to have ELM WD companions. These binaries might have formed tight or they were made tight due to stellar encounters within the cluster.  As the BLs evolve, it is quite possible that they engulf the ELM WDs in very close orbit, to become a relatively massive BSS. 

\par






\vspace{-2em}
\section{Conclusion}

The formation pathways of BSSs in OCs are more-or-less understood with the detection of WD companions to BSSs in clusters such as M67 and NGC 188. The details of the physics of MT, which are crucial to understand and predict the end products, are still missing. The detection of ELM WDs as companions to BSSs and BLs have confirmed MT pathway beyond any doubt. The detection of ELM WD companions to MS stars have opened up a new class of objects, known as blue lurkers, clearly challenge the basic definition of BSSs. Here we report the discovery of a second blue lurker in M67, with an ELM WD as companion. The fact that the fast rotating BSSs have relatively hotter WDs suggest that rotation could be used as a proxy to identify systems that have undergone stellar interactions. We also suggest that M67 has been forming the hot and luminous BSSs, as the young massive WDs detected by the UVIT require such progenitors. The BSS+WD system by UVIT in the globular cluster NGC 5466, points to the OC like environment in the outskirts of low density globular clusters.




\section*{Acknowledgements}
UVIT project is a result of collaboration between IIA, Bengaluru,  IUCAA,  Pune,  TIFR,  Mumbai,  several  centres  of ISRO, and CSA. This publication uses the data from the \textit{ASTROSAT} mission of the Indian Space Research  Organisation  (ISRO),  archived  at  the  Indian  Space  Science  Data Centre (ISSDC).

\vspace{-1em}


\bibliography{references}

\end{document}